\documentclass[prb,aps,twocolumn,showpacs,floats]{revtex4}
\usepackage{epsfig,bm,epsf,graphics}
\begin{document}
\title{Effect of Disorder in the Frustrated Ising FCC Antiferromagnet: Phase Diagram and Stretched Exponential Relaxation}
\author{V. Thanh Ngo$^{a}$, D. Tien Hoang$^{b}$, H. T. Diep$^{c}$\footnote{ Corresponding author, E-mail:diep@u-cergy.fr } and I. A. Campbell$^d$}
\address{$^a$ Institute of Physics, Vietnam Academy of Science and Technology\\
10 Daotan, Thule, Badinh, Hanoi, Vietnam\\
$^b$ High School for Gifted Students, Vinh University, \\
182 Leduan, Vinh city, Nghean province, Vietnam\\
$^c$ Laboratoire de Physique Th\'eorique et Mod\'elisation, Universit\'e de Cergy-Pontoise, CNRS UMR 8089\\
2, Avenue Adolphe Chauvin, 95302 Cergy-Pontoise Cedex, France\\
$^d$ Laboratoire Charles Coulomb, Universit\'e Montpellier II, 34095 Montpellier, France}

\begin{abstract}
We study the phase transition in a face-centered-cubic  antiferromagnet with Ising spins as a function of   the concentration $p$ of ferromagnetic bonds randomly introduced into the system. Such a model describes the spin-glass phase at strong bond disorder. Using the standard Monte Carlo
simulation and the powerful Wang-Landau flat-histogram method, we carry out in this work intensive simulations over the whole range of $p$. We show that the first-order transition disappears with a tiny amount of ferromagnetic bonds, namely $p\sim 0.01$, in agreement with theories and simulations on other 3D models. The  antiferromagnetic long-range order is also destroyed with a very small $p$ ($\simeq 5\%$).
With increasing $p$, the system changes into a spin glass and then to a ferromagnetic phase when $p>0.65$.
The phase diagram in the space ($T_c,p$) shows an asymmetry, unlike the case of  the $\pm J$ Ising spin glass on the simple cubic lattice.  We  calculate  the relaxation time around the spin-glass transition temperature and we show that the spin autocorrelation follows a stretched exponential relaxation law where the factor $b$ is equal to $\simeq 1/3$ at the transition as suggested by the percolation-based theory. This value is in agreement with experiments performed on various spin glasses and with Monte Carlo simulations on different SG models.
\end{abstract}

\maketitle

\section{Introduction}

Spin glasses \cite{Nishimori,Fischer,Mezard} (SG) have been a subject of intensive investigations for more than four decades. The main difficulties in such systems come from the combination of the frustration \cite{Diep2005} and the bond disorder.

One of the most interesting questions is the law of the relaxation time in disordered systems.  There is a large number of theories yielding various results since the first pioneering work by R. Kohlrausch which introduced phenomenologically in 1847 the so-called stretched exponential relaxation (SER) law. The reader is referred to the review by Phillips \cite{Phillips} for numerous examples of systems, not limited to SG, with  theoretical and experimental results.  We are interested in this paper in SG:  there have been several theoretical and numerical works dealing with the SER.
Ogielski \cite{Ogielski} has studied an Ising spin model of SG on a simple cubic lattice with a random nearest-neighbor $\pm J$ bond distribution by using intensive Monte Carlo (MC) simulations. Among his numerous results, he found that dynamic correlations follow the SER above the spin-glass transition temperature $T_g$ and a power law below $T_g$ for short and long times.  The SER is written as
\begin{equation}\label{SER}
q(t)=\overline{<S (0) S (t)>}=Ct^{-x}\exp (-\omega t^{b})
\end{equation}
where $S (t)$ is an Ising spin at the time $t$, $<...>$ indicates the thermal average, $\overline{<...>}$ denotes an average over disordered samples , $C$ a constant, $x$ and $b$ are temperature-dependent coefficients.  Ogielski found that $b$ is equal to $\simeq 1/3$ at $T_g$ and  it increases with increasing temperature ($T$). Below $T_g$,  he found that $q(t)$ follows the algebraic decay $Ct^{-x}$ where $x$ does not have the same temperature-dependence above and below $T_g$.
De Dominicis {\it et al.} \cite{DeDominicis} found by a calculation in the random-free-energy landscape that the relaxation to equilibrium follows a SER which depends on the choice of the transition probability between valleys in the free energy space.
There have been a number of other theories showing a SER behavior among which we can mention   the trap model\cite{Donsker}, the hierarchical model\cite{Palmer}, and the model of random walkers on dilute hypercubes of high dimensions\cite{Campbell85,Campbell87,Campbell93,Almeida}.  In the last model,
Almeida {\it et al.} \cite{Almeida} found, by a calculation using a random walk on the dilute hypercube, a SER near the percolation transition which is similar to the topology of spin configurations near the  transition in Ising SG. They found that $b$ is equal to $1/3$ at the percolation threshold $p_c$ and it increases with increasing $p$.  A recent MC simulation of the Sherrington-Kirkpatrick Ising SG model at and above $T_g$ shows a long-time SER  with $b$ equal to $\simeq 1/3$ at $T_g$\cite{Campbell2011}. Experimentally, let us mention a few data which show a SER with precise values of $b$. Early works on canonical SG Ag:Mn \cite{Chamberlin,Hoogerbeets} found a SER below $T_g$ with $b=1-n$ where $n\simeq 0.62$ for  $T/T_g$ from 0.5 to 0.8.  In an experiment on Cu$_{0.5}$Co$_{0.5}$Cl$_2$-FeCl$_3$ graphite bi-intercalation compound, Suzuki and Suzuki\cite{Suzuki} observed a $b$ value nearly equal to 0.3 at $T_g$.  Malinowski {\it et al.} \cite{Malinowski2011} found in Ni-doped La$_{1.85}$Sr$_{0.15}$CuO$_4$ a best fit of their time dependence of the relaxation with a SER.

Another point which motivates our present work is the effect of quenched bond randomness in systems having a first-order transition.  Note that in a pure system with a second-order transition, the effect of a small amount of bond disorder is understood according to the Harris criterion \cite{Harris}. In systems with a first-order transition, this question has been examined since 1989 by a number of important investigations mostly in two dimensions (2D)\cite{Imry,Aizenman,Hui,Cardy}, inspired from the pioneering work of Imry and Ma on the 2D Random-Field Ising Model \cite{ImryMa}.  These works have shown  that a small quenched bond disorder suffices to make disappear the latent heat in a pure 2D system having a first-order transition. In particular, Hui and Berker\cite{Hui}  have demonstrated this by looking at the Blume-Emery-Griffiths (BEG) spin-1 model in 2D,  with a renormalization group (RG) argument. Cardy and Jacobsen \cite{Cardy} have illustrated this with the $q=8$ Potts model by the use of finite-size scaling and conformal invariance. There is however a contradiction between different researchers concerning the universality of the resulting second-order transition upon introducing a small disorder: some authors \cite{Hui,Cardy,Cardy87,Ludwig,Dotsenko} claimed that this should belong to a new random fixed point for $q>2$, while others favored the 2D Ising universality class \cite{ChenLandau}. In 3D, Falicov and Berker \cite{Berker1996} have shown by a RG calculation that the tricritical point in the pure BEG model is replaced, when bond randomness is introduced,  by a segment of second-order transitions bounded at one end by a multicritical point and at the other end by a new random tricritical point.  Interestingly, they have shown that there exists a threshold of randomness beyond which the random tricritical point goes down to $T=0$. In a recent paper, Malakis {\it et al.}\cite{Berker2012} have carried out a MC study on the 3D BEG model on a simple cubic lattice. They showed that the tricritical point moves in the direction to reduce the first-order line in the phase diagram and
the so-become second-order transition segment has the critical behavior of the 3D random Ising model.
In addition, Fernandez {\it et al.}\cite{Fernandez2012} have also shown by microcanonical MC simulations that the 3D site-dilute 4- and 8-state Potts models confirm the conjecture of Cardy and Jacobsen \cite{Cardy} in 3D.  For quantum spin systems, Greenblatt {\it et al.}\cite{Aizenman2009} showed rigorously the disappearance of first-order transition with randomness in 2D and in dimension $d\leq 4$, just as for the classical case \cite{Aizenman}.  Note that in the case where the pure model has a second-order transition such as when $q=4$, Domany and Wiseman \cite{Domany} have shown that a random quenched bond disorder changes the second-order $q=4$ Potts universality into a 2D pure Ising one.

The present paper addresses the two points mentioned above: (i) to verify the conjecture of Cardy and Jacobsen \cite{Cardy} on the disappearance of the latent heat in the case of a 3D first-order transition when a quenched bond disorder is introduced (see Fig. 1 of their paper), and to see if the result found for the 3D BEG model\cite{Berker1996,Berker2012} applies to other models such as the one studied in the present paper (ii) to study dynamic correlations of the spin-glass phase created at a strong bond disorder of our model to verify that the results of Ogielski are model-independent.   To carry out these purposes, we consider the pure Ising face-centered cubic (FCC) antiferromagnet with nearest-neighbor (NN) interaction.  This model is fully frustrated since the lattice is composed of equilateral triangles with antiferromagnetic interaction \cite{Diep2005}. It shows a very strong first-order transition \cite{Phani,Polgreen,Styer,DiepAlloy1986}.

The paper is organized as follows. Section II is devoted to a description of our model and the simulation method. Results are shown and discussed in section III. Concluding remarks are given in section IV.

\section{Model and Wang-Landau algorithm}

\begin{figure}
\centerline{\epsfig{file=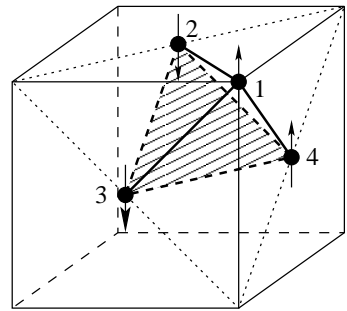,width=1.8in}} \caption{The FCC cell with four sublattices.  An elementary tetrahedron is shown.} \label{fig:fcc-cell}
\end{figure}

We consider the FCC lattice with Ising spins of magnitude $S=1$. The Hamiltonian is given by
\begin{equation}
{\cal H} = -\sum_{<i,j>} J_{ij}S_i\ S_j
\end{equation}
where $S_i$ is the Ising spin at the lattice site $i$, $\sum_{(i,j)}$ is made
over the NN spin pairs $S_i$ and $S_j$ with  interaction $J_{ij}$. Hereafter we suppose that $J_{ij}=-J$ ($J>0$) for antiferromagnetic bonds and $J_{ij}=J$ for ferromagnetic bonds. The  lattice is composed of $L^3$
FCC-cells each with four spins ($L$ being the number of cells in each direction). The four sublattices are shown in Fig.~\ref{fig:fcc-cell}. The total number of spins of the lattice is $N=4L^3$. We use the periodic boundary conditions.  The number of NN bonds per spin is $12$, thus the total number of bond of the system is $N_b=12\times N/2 = 24 L^3$.  Hereafter, the energy is measured in the unit of $|J|=1$ and $T$ in the unit of $|J|/k_B=1$.

We start with the pure antiferromagnetic state.  For this case, it is well-known that the bulk FCC antiferromagnet with Ising spins shows a very strong first-order transition \cite{Phani,Polgreen,Styer,DiepAlloy1986}.
 The phase transition in the Heisenberg model also shows a first-order character \cite{Diep1989fcc,Gvoz}.  Other similar
frustrated antiferromagnets such as the HCP antiferromagnet show
the same behavior \cite{Diep1992hcp,Hoang2012}.

To create a spin glass, we introduce a number of ferromagnetic bonds $N_b^F$ in a random manner into the pure antiferromagnetic state.  The ferromagnetic bond concentration $p$ is thus  $p=N_b^F/N_b$. The phase transition behavior of the system depends on the parameter $p$ ($0\le p\le 1$), i.e. it changes from pure antiferromagnetic phase at $p=0$ to pure ferromagnetic phase at $p=1$, passing through the spin-glass phase.

In order to investigate the nature of the phase transition and the relaxation behavior at various $p$, we use the standard MC method and the Wang-Landau technique of simulation. Wang and Landau\cite{WL1} have proposed a MC algorithm for classical statistical models which allowed to study systems with difficultly accessed microscopic states. In particular, it allows us to detect  with efficiency weak first-order transitions\cite{Ngo08,Diep2008}. The algorithm uses a random walk in the energy space to get an accurate estimate for the density of states (DOS)  $g(E)$ which is defined as the number of spin configurations for any given $E$. This method is based on the fact that a flat energy histogram $H(E)$ is produced if the probability for the transition to a state of energy $E$ is proportional to $g(E)^{-1}$.
We summarize how this algorithm is implied here. At the beginning of the simulation, the density of states $g(E)$ is unknown so all densities are set to unity, $g(E)=1$.
We begin a random walk in energy space $(E)$ by choosing a site randomly and flipping its spin with a transition probability
\begin{equation}
p(E\rightarrow E')=\min\left[g(E)/g(E'),1\right],
\label{eq:wlprob}
\end{equation}
where $E$ is the energy of the current state and $E'$ is the energy of the proposed new state.
Each time an energy level $E$ is visited, the DOS is modified by a modification factor $f>0$ whether the spin is flipped or not, i.e. $g(E)\rightarrow g(E)f$.
  At the beginning of the random walk, the modification factor $f$ can be as large as $e^1\simeq 2.7182818$. A histogram $H(E)$ records the number of times a state of energy $E$ is visited. Each time the energy histogram satisfies a certain ``flatness" criterion, the  histogram $H(E)$ is then reset to zero, and the modification factor is reduced, typically to the square root of the previous factor, to produce a finer estimate of $g(E)$. The reduction process of the modification factor $f$ is repeated several times until a final value $f_{\mathrm{final}}$ which is close enough to one. The histogram is considered as flat if
\begin{equation}
H(E)\ge x\%\ \langle H(E)\rangle
\label{eq:wlflat}
\end{equation}
for all energies, where $x\%$ is chosen between $90\%$ and $95\%$
and $\langle H(E)\rangle$ is the average histogram.

The thermal average of a thermodynamic quantity $A$ can be evaluated by\cite{WL1,brown}
\[\langle A\rangle_T =\frac{1}{Z}\sum_E g(E)A \exp(-E/k_BT),\]
where $Z$ is the partition function defined by
\begin{equation}
Z =\sum_E g(E)\exp(-E/k_BT)
\label{eq:partfunc}
\end{equation}
The canonical distribution at any $T$ can be calculated simply by
\begin{equation}
P(E,T) =\frac{1}{Z}g(E)\exp(-E/k_BT)
\label{eq:pe}
\end{equation}

In this work, we consider an energy range of interest\cite{Schulz,Malakis}
$(E_{\min},E_{\max})$. We divide this energy range into $R$ subintervals, the minimum energy of each subinterval is $E^i_{\min}$ for $i=1,2,...,R$, and the maximum of the subinterval $i$ is $E^i_{\max}=E^{i+1}_{\min}+2\Delta E$,
where $\Delta E$ can be chosen large enough for a smooth boundary between two subintervals. The WL
algorithm is used to calculate the relative DOS of each subinterval $(E^i_{\min},E^i_{\max})$ with the
modification factor $f_\mathrm{final}=\exp(10^{-9})$ and flatness criterion $x\%=95\%$.
We reject the suggested spin flip and do not update $g(E)$ and the energy histogram $H(E)$ of
the current energy level $E$ if the spin-flip trial would result in an energy outside the energy segment.
The DOS of the whole range is obtained by joining the DOS of each
subinterval $(E^i_{\min}+\Delta E,E^i_{\max}-\Delta E)$.

\section{Results}

We performed runs with different random quenched bond disorders at each given $p$ for bond-configuration averaging. The Edwards-Anderson order parameter, the magnetization (and sublattice magnetization) and the susceptibility are obtained by standard MC simulations with the equilibration time of $t_e=2\times 10^6$ MC steps/spin and the averaging time of $t_a=4\times 10^6$ MC steps/spin. The energy, specific heat and energy histograms are obtained by the WL technique in order to detect with precision the latent heat \cite{Ngo08,Diep2008}.

\subsection{Nature of the phase transition}
\begin{figure}[!hbt]
\centerline{\epsfig{file=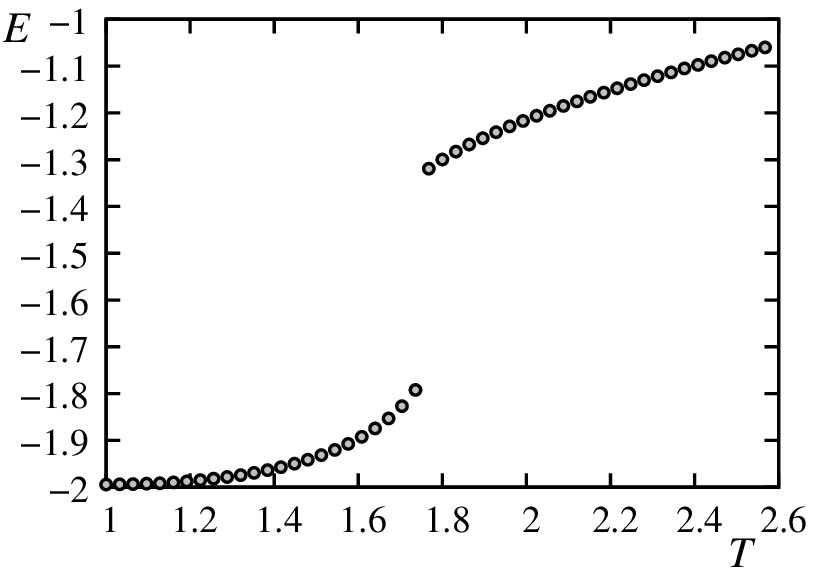,width=6cm}}
\centerline{\epsfig{file=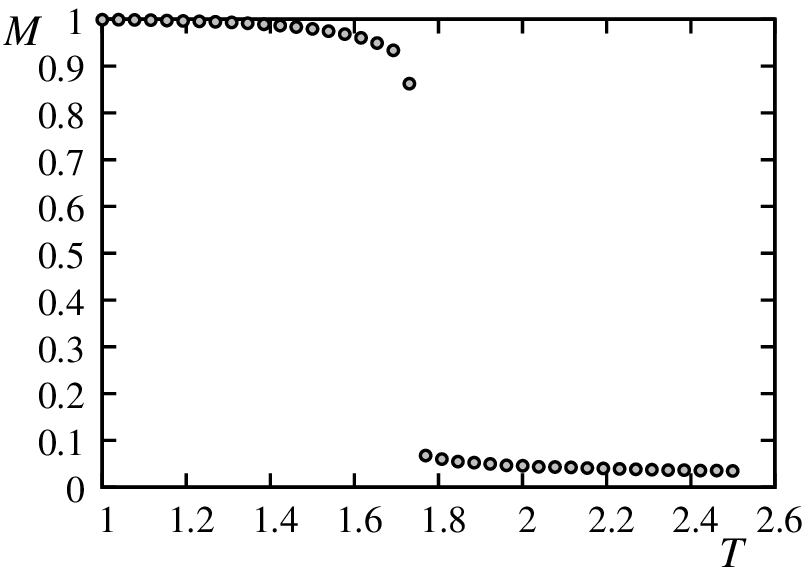,width=6cm}}
\caption{Energy (top) and sublattice magnetization (bottom) versus  $T$ for $p=0$ and $L=12$. This is the case of pure FCC antiferromagnet.}
\label{fig:ep0000}
\end{figure}

\begin{figure}[!hbt]
\centerline{\epsfig{file=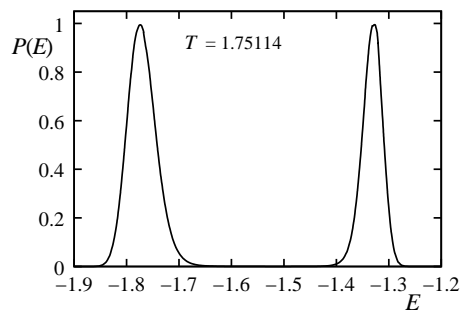,width=6cm}} \caption{Energy histograms as a function of energy at $T=1.75114$  with $p=0$ and $L=12$.}
\label{fig:pep0000}
\end{figure}

In Fig.~\ref{fig:ep0000}, we show the case of $p=0$ where the energy and the sublattice magnetization versus $T$ undergo a well-known first-order transition\cite{Phani,Polgreen,Styer,DiepAlloy1986} with a discontinuity at the transition temperature $T_c$.  Figure \ref{fig:pep0000} shows a double peak of the energy distribution at $T_c$ which confirms the first-order behavior.


\begin{figure}
\centerline{\epsfig{file=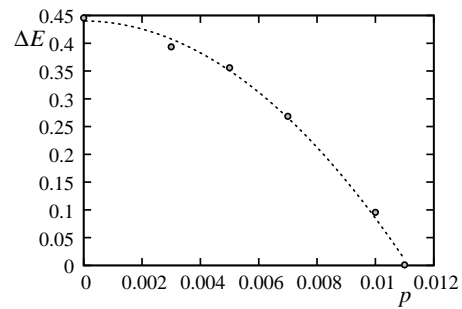,width=6cm}} \caption{The latent heat versus $p$.}
\label{fig:de}
\end{figure}

The latent heat $\Delta E$ is defined by the energy separation  of the two peaks in the energy distribution.  When the transition is of second order, the energy is continuous, namely $\Delta E=0$.  Using the WL technique which is very efficient to detect first-order transitions, we calculate the energy histogram with various small values of $p$.
 We show in Fig.~\ref{fig:de} the latent heat versus $p$. As seen,  $\Delta E$ is not zero for $p \leq 0.011$. For $p > 0.011$, the phase transition is continuous.  The disappearance of the latent heat for such a small amount of bond disorder is striking. This result is in agreement with the RG result on the 3D BEG model\cite{Berker1996}, with the conjecture of Cardy and Jacobsen\cite{Cardy} for the 3D case (see Fig. 1 of their paper), and with predictions of other earlier works \cite{Imry,Aizenman,Hui}.  To our knowledge, the present work is the first verification in 3D of the disappearance of the latent heat with a tiny amount of bond disorder.  There is thus a tricritical point at $p_c\simeq 0.011$ beyond which the transition is of second order. This result is  in agreement with MC simulations on the 3D BEG model\cite{Berker2012} and site-dilute Potts model\cite{Fernandez2012}.   An interesting question arises on the nature of the second-order transition coming from the ex first-order transition. Malakis {\it et al.}\cite{Berker2012} have shown that the critical behavior belongs to the 3D random Ising model in violation of the universality of the whole second-order line. To verify this result on our present FCC frustrated model, we need intensive MC simulations with high-performance techniques such as multiple histograms. This formidable task is not the scope of the present paper.  Note that the change of the nature of the phase transition with randomness occurs also in the so-called fully frustrated simple cubic lattice with Ising spins\cite{Ngo2011,Berker1984}: it was found by MC simulations that a very small amount of bond dilution changes the transition from a weak first-order transition into a SG transition \cite{Campbell1995}.

For the case of pure ferromagnet, i.e. $p=1$, the transition is a second-order one as expected for a 3D Ising ferromagnet.


\subsection{Phase diagram}

Let us consider intermediate values of $p$ where the spin-glass phase is expected.
The spin-glass phase is determined by the Edwards-Anderson freezing order parameter defined as \cite{EdwardsAnderson,BinderYoung}
\begin{equation}\label{qea}
Q =\frac{1}{N} \frac{1}{t_a}\sum_{i}^{N}\left|\sum_{t=t_e}^{t_e+t_a} S_i(t)\right|
\end{equation}
The averaging time $t_a$ is taken as long as possible (at least several millions of MC steps per spin), after an equivalent equilibration time $t_e$.  In the calculation of $Q$, we follow each spin during the time $t_a$ while calculating its time-averaged value. At the end, we take an average over all spins. So, $Q$ expresses the degree of spin freezing, independent of spin configuration.

The magnetization is defined as
\begin{equation}\label{mag}
M=\frac{1}{N}\frac{1}{t_a}\sum_{t=t_e}^{t_e+t_a}\sum_i^{N} S_i(t)
\end{equation}
Here, $M$ is  averaged over all spins before the time averaging.
In a  state with randomly frozen spins, the magnetization $M$ is zero  so that it cannot distinguish a frozen phase from the paramagnetic phase.

\begin{figure}
\centerline{\epsfig{file=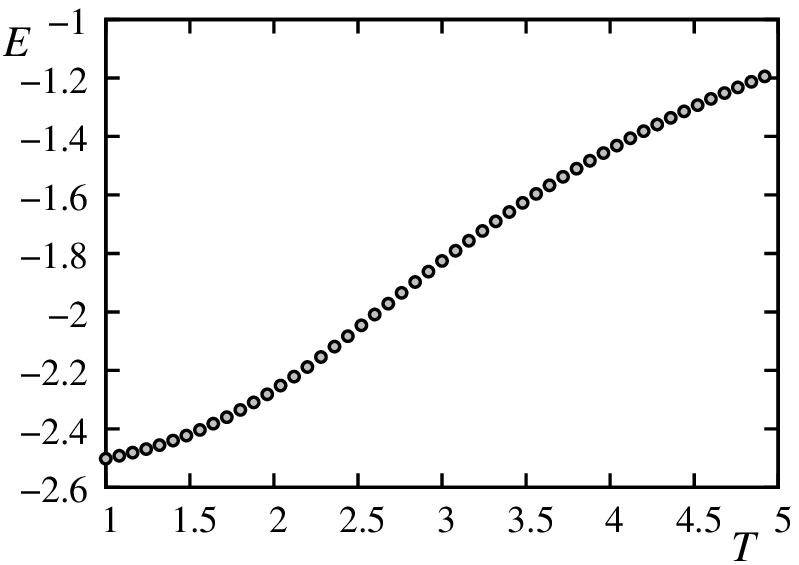,width=6cm}}
 \centerline{\epsfig{file=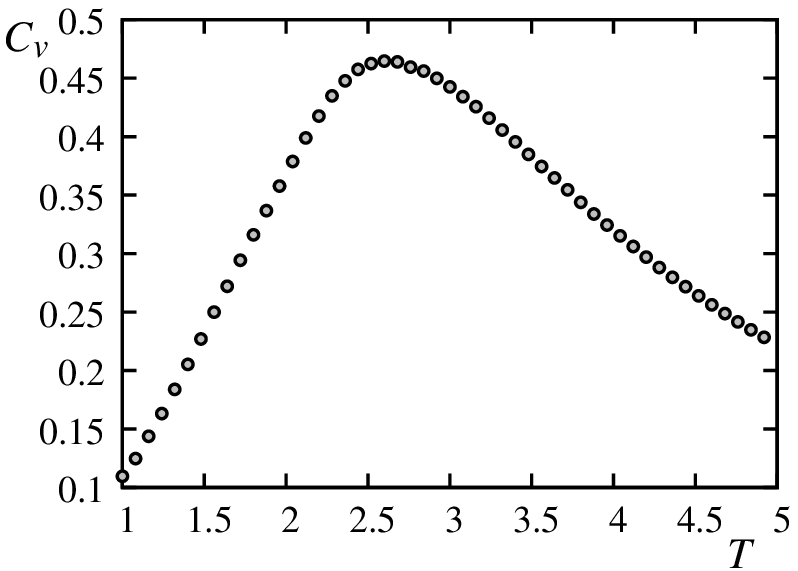,width=6cm}}
 \caption{Energy (top) and specific heat (bottom) versus $T$ for $p=0.5$ and $L=12$.}
\label{fig:ep0500}
\end{figure}

We show now in Fig. \ref{fig:ep0500}  the energy and the specific heat versus $T$ for $p = 0.5$ where the bond disorder is strongest.

\begin{figure}
\centerline{\epsfig{file=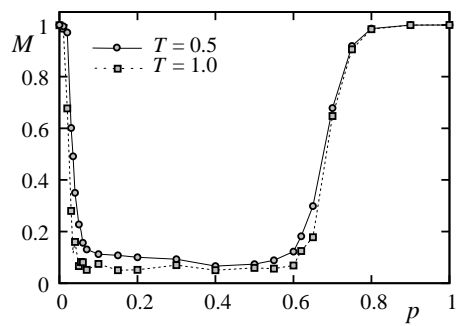,width=6cm}}
 \caption{Magnetization (or sublattice magnetization for antiferromagnetic side)
 $M$ versus $p$ at $T=0.5$ and 1.   }
\label{fig:mpt}
\end{figure}

The magnetization $M$ versus $p$ at  $T=0.5$ and $T=1.0$  is shown in Fig.~\ref{fig:mpt}. Note that for the antiferromagnetic side (small $p$) we use in that figure the sublattice magnetization.  If a spin-glass phase is defined as the one where the magnetization (or sublattice magnetization) is zero but not $Q$ at low $T$, then we find that the spin-glass phase exists in the range $0.055 < p < 0.62$.

\begin{figure}
\centerline{\epsfig{file=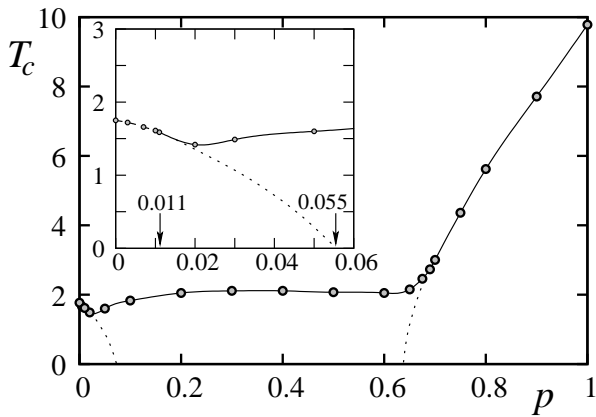,width=8cm}} \caption{$T_c$ versus the ferromagnetic bond concentration $p$ with $L=12$. The inset shows an enlarged scale of the small $p$ region: the discontinued  line at $p<0.011$ indicates the first-order transition line. The dotted lines are the reentrant lines in the regions $0.02<p<0.055$ and $0.65<p<0.68$ (data taken from Fig. \ref{reentrance}).}
\label{fig:tcp}
\end{figure}

\begin{figure}
\centerline{\epsfig{file=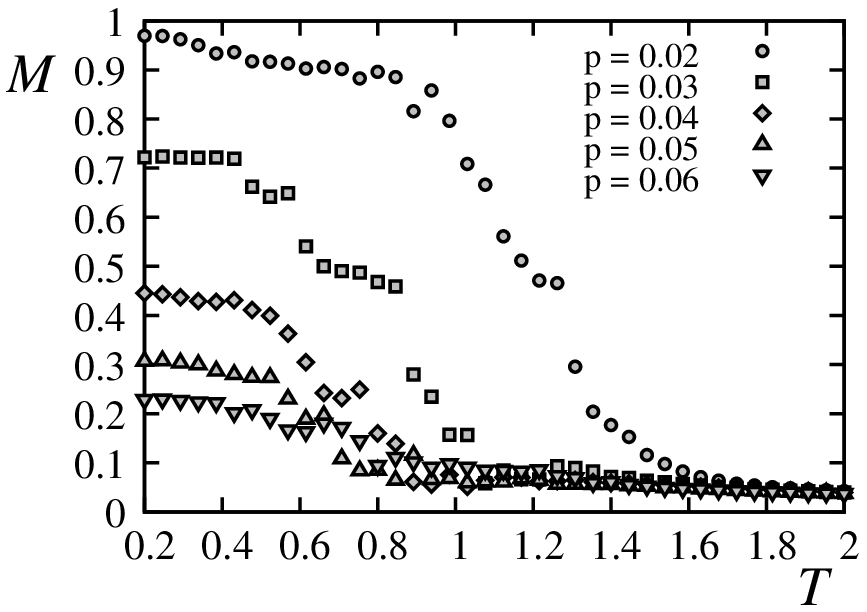,width=6cm}}
\centerline{\epsfig{file=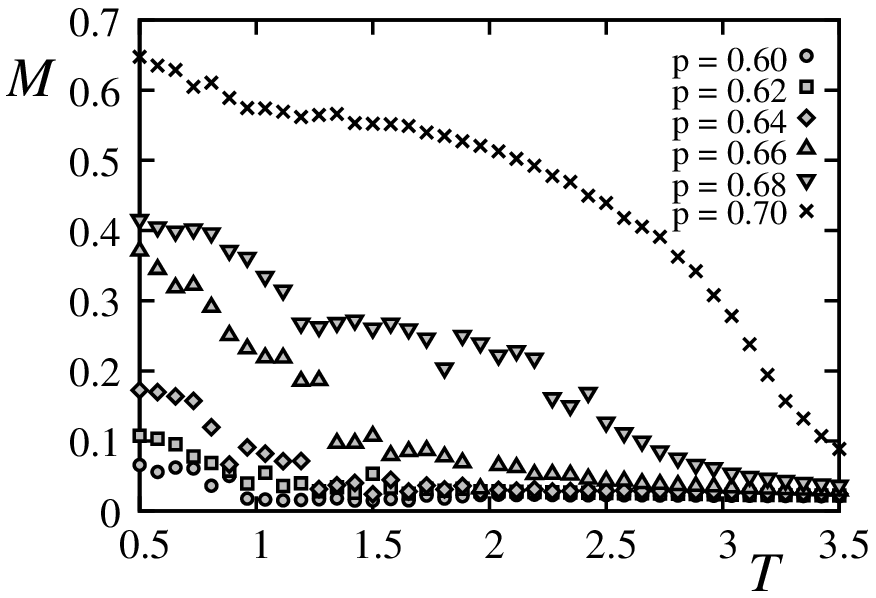,width=6cm}}
\caption{Sublattice magnetization in the region $0.02\leq p \leq 0.06$ and magnetization in the region $0.60\leq p \leq 0.70$, versus $T$. The temperatures at which the sublattice magnetization (AF side) or the magnetization (F side) go to $\simeq 0$ below $T_c$ are shown  by the dotted lines in Fig. \ref{fig:tcp}. Those lines indicate the limit of the reentrant phases. }
\label{reentrance}
\end{figure}

The phase diagram $(T_c,p)$ is shown in Fig.~\ref{fig:tcp} where $T_c$ is obtained at each $p$ when $Q$ goes down to zero.  Note that the maxima of the specific heat $C_v$ and the susceptibility $\chi$ are located at a temperature higher than $T_c$.  For example, at $p=0.5$, $Q=0$ at $T=T_c= 2\pm 0.03$ while the peaks of $C$ and $\chi$ are at $T\simeq 2.6$.
The fact that the maxima of the specific heat and the susceptibility are located at a
temperature higher than $T_c$ does not result from a finite-size effect. These maxima are
due to the change of the correlation nature in the paramagnetic phase. A very detailed discussion on this
 phenomenon has been given by Ogielski \cite{Ogielski} and by Binder and Young \cite{BinderYoung}. Briefly, above the maxima of $C_v$ and $\chi$  the system is in a "normal" paramagnetic regime (short-range correlation, short relaxation times,...). At the maximum of $C_v$ (and $\chi$), the SG behavior sets in with dramatic rapid increase of SG correlation length and correlation times as well as the nonlinear SG susceptibility. However, only at the SG transition temperature $T_c$ that these quantities diverge.
There  are  several striking features shown in Fig. \ref{fig:tcp}:

i) The phase diagram is not symmetric with respect to $p$.   This is due to the fact that the FCC lattice is fully frustrated at $p=0$ (antiferromagnetic) and non-frustrated at $p=1$.  Note that the case where the pure system is non-frustrated such as the simple cubic ferro- or antiferromagnets (bipartite lattice), the phase diagram $(T_c,p)$ is symmetric because the system is invariant with respect to the local transformation of every spin pair ($J_{ij}\rightarrow -J_{ij}, S_j\rightarrow -S_j)$.

ii) The first-order transition line terminates at $p=0.011$ (see inset).  As said earlier, in the present 3D model, such a tiny bond randomness suffices to change the first-order transition into a continuous transition.

iii) The long-range antiferromagnetic ordering is destroyed for $p \geq 0.055$ (we examined the sublattice magnetization $M$ down to $T\simeq 0$).
  The fact that such a very small $p$ suffices to destroy the long-range  antiferromagnetic order is not a real surprise if we examine the nature of the long-range ordering in the antiferromagnetic limit: it is known that the ground-state (GS) spin configuration of the FCC antiferromagnet is a stacking of  corner-sharing tetrahedra. Each tetrahedron has six choices of two spins up and two spins down on its corners (see Fig. \ref{fig:fcc-cell}).  Such a stacking gives rise to an infinite number of  GS configurations which are random with no long-range ordering except for 6 configurations composed of 3 configurations by stacking of up-spin planes and down-spin planes, alternately in $x$, or $y$ or $z$ direction and 3 configurations due to the global reversal of all spins.  When $T$ increases from zero, the system selects these 6 long-range GS configurations. This phenomenon is known as the selection of  ``order by disorder" \cite{Bidaux,Henley} which has been numerically verified in a large number of systems \cite{Diep1989fcc,Diep1992hcp,Pinettes1998,Hoang2012}.  The selection of the long-range order is due to the entropy term which makes an extremely small difference with the other nearly-degenerate random configurations.  It is not therefore surprising that the introduction of even a very small amount of ferromagnetic bonds breaks immediately the long-range order.

iv) It is interesting to note that we have observed two regions of reentrant phase at $0.02<p<0.055$ and $0.65<p<0.68$, bounded at the upper limit by the dotted lines.    In these regions, at a given $p$ the magnetization is small at low $T$ and becomes zero when crossing the dotted line,  below the transition line determined by the vanishing Edwards-Anderson order parameter $Q$.  The dotted lines are determined  using the temperatures at which the magnetization (or sublattice magnetization for AF side) vanish for different $p$ as shown in Fig. \ref{reentrance}. Note that, depending on the system, there are several complex mechanisms leading to reentrant phases as seen in exactly solved Ising models \cite{BerkerNishimori,DiepGiacomini}.

\subsection{Stretched exponential relaxation}

To study the relaxation of the freezing order parameter in the ``steady" regime (after equilibration time $t_e$), let us define the following quantities
\begin{eqnarray}
Q_{\Delta t}(t) &=&\frac{1}{N}\frac{1}{\Delta t} \sum_{i}^{N}\left|\sum_{t'=t}^{t+\Delta t} S_i(t')\right|,\label{eq:o}\\
\overline{Q_{\Delta t}} &=&\frac{1}{N_I} \sum_{t} Q_{\Delta t} (t), \label{eq:mq}
\end{eqnarray}
where $\sum_{t'=t}^{t+\Delta t}$ indicates a sum over $\Delta t$ MC steps per spin starting from $t$ and $\sum_{i}^{N}$ indicates a sum over the system sites. In Eq.~(\ref{eq:mq}), $\overline{Q_{\Delta t}}$  is the average of $Q_{\Delta t}$ calculated over all successive intervals of $\Delta t$ steps until the end of the total run time $t_a$.  $\overline{Q_{\Delta t}}$  is thus averaged with $N_I=t_a/\Delta t=10^4$ intervals, taking $t_a=10^7$ and $\Delta t=1000$, for instance.
Note that $\Delta t$ is nothing but the so-called ``waiting time" when measuring the system relaxation.

We emphasize that $Q_{\Delta t}(t) $ is in fact the autocorrelation function. This is seen by writing the autocorrelation of the spin $S_i$ after a waiting time $\Delta t$ as
\begin{eqnarray}
&&q_{i,\Delta t}(t)=\frac{1}{\Delta t}\left |\sum_{t'=t}^{t+\Delta t} S_i (t) S_i (t')\right |\nonumber\\
&&=\frac{1}{\Delta t}\left |\pm \sum_{t'=t}^{t+\Delta t} S_i (t')\right |\ \ \mbox{(since $S_i(t)=\pm 1$}) \nonumber\\
&&=\frac{1}{\Delta t}\left |\sum_{t'=t}^{t+\Delta t} S_i (t')\right |\nonumber
\end{eqnarray}
from which we obtain $Q_{\Delta t}(t) $ by summing over all spins.
When $\Delta t=t_a$ we have $\overline{Q_{\Delta t}} \equiv Q$.

\begin{figure}
\centerline{\epsfig{file=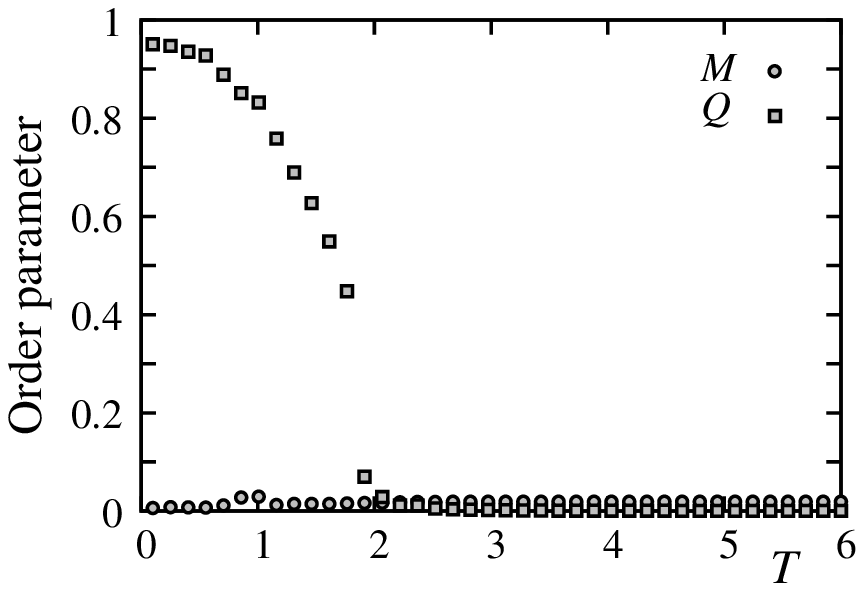,width=6cm}}
\centerline{\epsfig{file=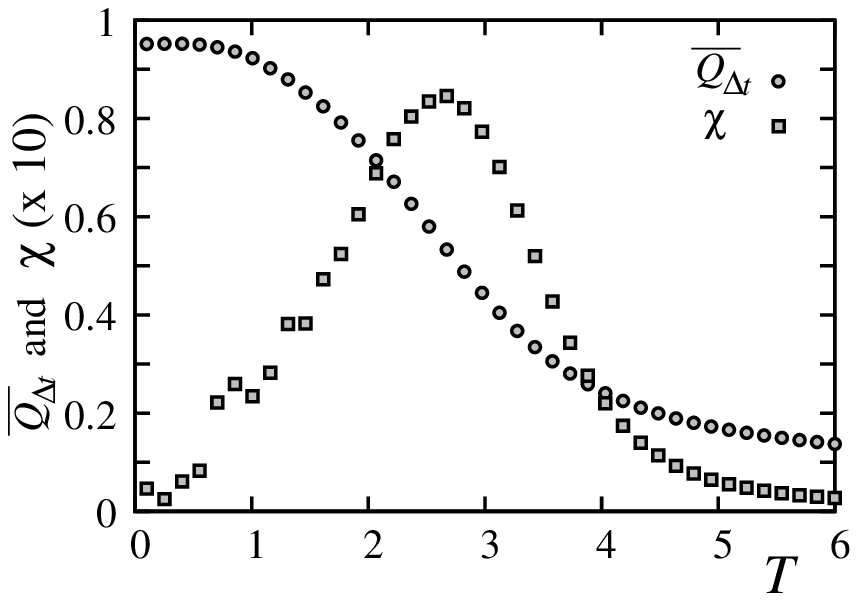,width=6cm}}
\caption{Top: Order parameter $Q$ and  magnetization $M$ vs $T$ for $p=0.5$, with averaging time $t_a=2\times 10^6$.
Bottom: $\overline{Q_{\Delta t}}$  and $\chi =N(\langle \overline{Q_{\Delta t}}^2\rangle-\langle \overline{Q_{\Delta t}}\rangle^2)/T$, versus $T$ for $p=0.5$, with $\Delta t=30000$. See text for comments.}
\label{fig:xp0500}
\end{figure}

As shown in Fig.~\ref{fig:xp0500}, the magnetization $M$ is  zero at any $T$ for $p=0.5$. Therefore, it is impossible to calculate the susceptibility defined by the variance of $M$. However, we can do that for $\overline{Q_{\Delta t}}$: $\chi= N(\langle \overline{Q_{\Delta t}})^2\rangle-\langle \overline{Q_{\Delta t}}\rangle^2)/T$. Of course, the result of $\chi$ will depend on  $\Delta t$. We show an example of $\chi$ for $\Delta t=30000$ in Fig.~\ref{fig:xp0500}.  As will be seen below, this waiting time is much longer than the relaxation time at $T$ higher than, but not too close to,  $T_c\simeq 2$. Therefore, $\chi$ shown in Fig.~\ref{fig:xp0500} is considered as time-independent in this temperature range with its peak at $T\simeq 2.6$.

\begin{figure}[h!]
\centerline{\epsfig{file=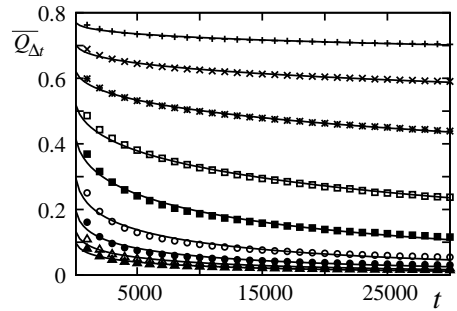,width=6cm}} \caption{$\overline{Q_{\Delta t}}(t)$ vs $t$ (in unit of MC step) at $T=1.5,\ 1.7,\dots\ 3.3$ (from top to bottom), for  $p=0.5$.}
\label{fig:rtp05}
\end{figure}

We use the following  SER function defined by
\begin{equation}
\label{eq:srf}
\overline{Q_{t}}=A\exp\left[-(t/\tau)^b\right]+a,
\end{equation}
where $t$ is the waiting time which is $\Delta t$ in our definition given above, $b$ the SER exponent, $A$ a temperature-dependent constant, and $\tau$  the SER relaxation time.  Note that this definition, without the constant $a$, has been used by most of previous authors \cite{DeDominicis,Almeida,Campbell2011,Suzuki,Malinowski2011}. They did not use the pre-factor $t^{-x}$  as in Eq. (\ref{SER}).  We have introduced $a$ in order to use the infinite-time limit for the fitting  because we have taken  $t$ from 1000 to 30,000 in the simulation which are rather short. At the infinite-time limit, $a$ is zero for $T\gg T_c$, and $a=Q$ for $T<T_c$.
We can assimilate   $Q$ calculated during several millions of MC steps to $\overline {Q_{t=\infty}}$.  Note that doubling the averaging time $t_a$ reduces the fluctuations of $\overline {Q_{t_a}}$ but does not significantly change its mean value at least up to the fifth digit, while the exponential term is much smaller, of the order of ${\mbox {e}}^{-10}$.  In our simulations, we take the equilibration time $t_e=4\times 10^6$ MCs/spin and the averaging time $t_a=2\times 10^6$ MCs/spin.
We plot in Fig.~\ref{fig:rtp05} the autocorrelation $\overline{Q_{\Delta t}}(t)$ for $p=0.5$ as a function of $t$ at $T=1.5,\ 1.7,\ \dots,\ 3.3$ obtained from the simulation.  Fitting these curves with the formula (\ref{eq:srf}) and using $Q_{t=t_a\simeq \infty}$,  we get the  parameters $A$, $a$ and $b$ and  $\tau$ which are shown in Tab.~\ref{tabab}.
Note that, the SER function in Ref.~\onlinecite{Campbell2011,Malinowski2011} was studied only for $T >T_c$.

\begin{table}
  \centering
\begin{tabular}{|c|c|c|c|c|}
\hline $T$ & $\tau$ & $b$ & $a$ & $A$ \\
\cline{1-5}
1.5 & $1.2292\times 10^7$ & 0.101793 & 0.246619 & 0.753381 \\
1.7 & $3.4434\times 10^6$ & 0.159285 & 0.150010 & 0.704799 \\
1.9 &   136080.  &  0.256315 &   0.094107 &   0.644986 \\
2.1  &  52089.4 &   0.396416  &  0.062483  &  0.533433 \\
2.3  &  12637.7  &  0.497942 &   0.043070  &  0.345517 \\
2.5  &  7525.13 &   0.565996  &  0.031611 &   0.207143 \\
2.7  &  4050.33 &   0.621610 &   0.024736 &   0.150738 \\
2.9 &   3839.15  &  0.652591  &  0.019274 &   0.100338 \\
3.1 &   3812.93  &  0.686200 &   0.015024 &   0.071225 \\
3.3  &  3790.41  &  0.708795  &  0.012179  &  0.054192 \\
\hline \end{tabular}
  \caption{Values of $a$, $b$, $A$, and $\tau$ at several  $T$ for $p=0.5$, $L=12$.}\label{tabab}
\end{table}

\begin{figure}
\centerline{\epsfig{file=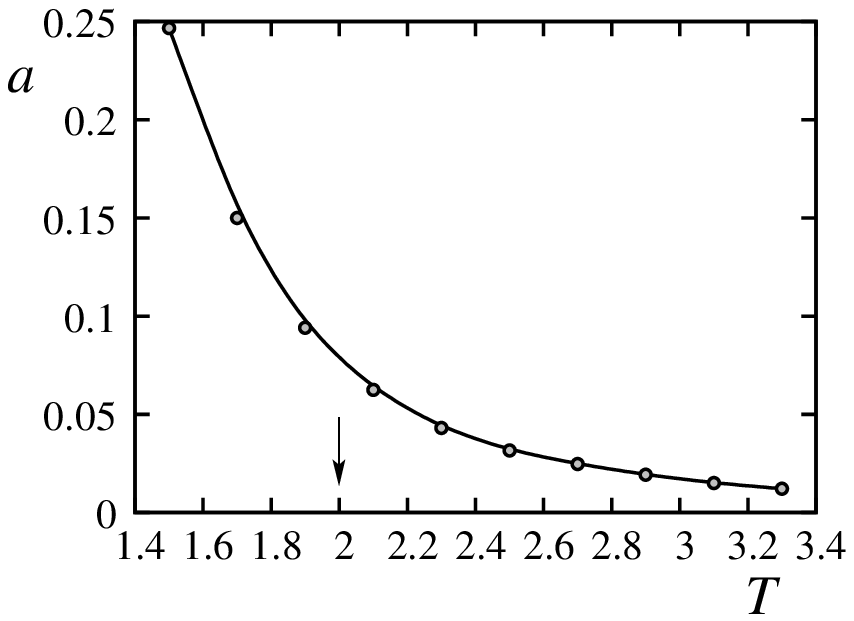,width=6cm}}
\centerline{\epsfig{file=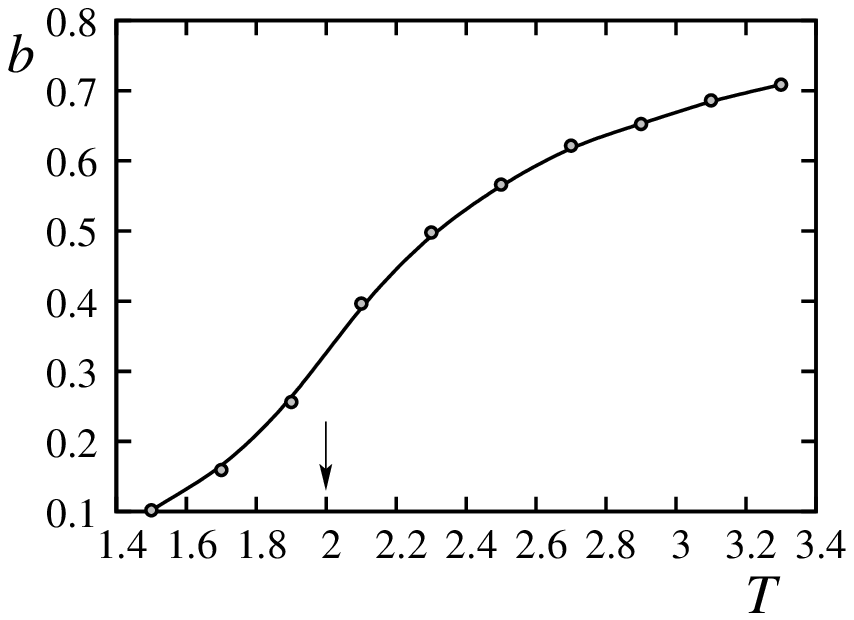,width=6cm}}
\centerline{\epsfig{file=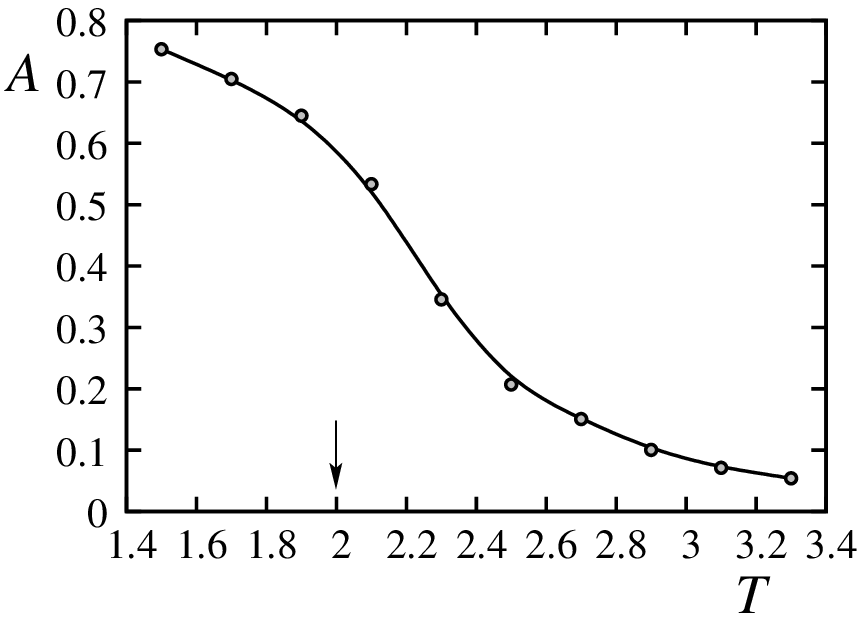,width=6cm}}
\caption{Parameters $a$ (top) and $b$ (middle), and $A$ (bottom) as  functions of $T$ for $p=0.5$, $L=12$. The arrow indicates the transition temperature. See text for comments.}
\label{fig:aot}
\end{figure}

\begin{figure}
\centerline{\epsfig{file=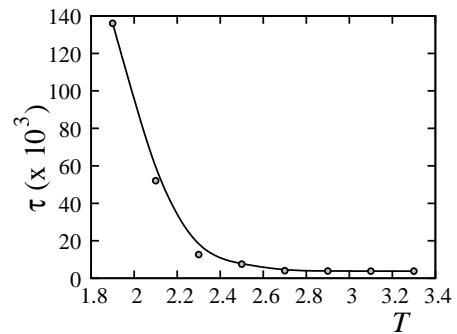,width=6cm}}
\caption{SER relaxation time $\tau$ as a function of $T$ for $p=0.5$, $L=12$.}
\label{fig:tauot}
\end{figure}

The dependence of  $a$, $b$, $A$ and $\tau$ on $T$ is shown in Figs. \ref{fig:aot} and \ref{fig:tauot}.
Several remarks are in order:

i) $a$ decreases with increasing $T$. However  it goes to zero only far from $T_c$, namely deep inside the paramagnetic phase. The residue value after $T_c$ is due to the finite size effect and to short-range correlations which exist at $T \gtrsim T_c$.

ii) $A$ changes the curvature at $T_c\simeq 2$. If this coefficient is related to the pre-factor $t^{-x}$ of Ogielski\cite{Ogielski}, then the change of curvature of $x$ at the glass transition temperature that he observed may be due to the same physical origin.

iii) $b(T)$ changes the curvature at $T_c\simeq 2$ where it is equal to $\simeq  0.33$.   Note that the determination of $b$ at a given $T$ does not depend on $T_c$, unlike the critical exponents.  The value $b=1/3$ at $T_c$ is what predicted by theories\cite{DeDominicis,Campbell85,Almeida}. It is also what was found numerically for different models of Ising SG: $\pm J$ on simple cubic lattice\cite{Ogielski} and Sherrington-Kirkpatrick model\cite{Campbell2011}.  This value is also close to what is experimentally found in various types of SG, not necessarily corresponding to Ising models \cite{Chamberlin,Hoogerbeets,Suzuki,Malinowski2011}. We believe that at least at $T_c$, $b=1/3$ is universal.
The value of $b$ increases with increasing $T$.  Note however that we did not go very far into the paramagnetic phase ($T_c<T<3.3$) so that $b$ is far from the expected value $b=1$ for the true paramagnetic phase ($T\gg T_c)$, but the tendency towards 1 is clearly seen in the curve of $b$ of Fig. \ref{fig:aot}.

iv) The SER relaxation time  $\tau$  becomes extremely large at $T\simeq 2$ as expected from the critical slowing down.


\begin{figure}
\centerline{\epsfig{file=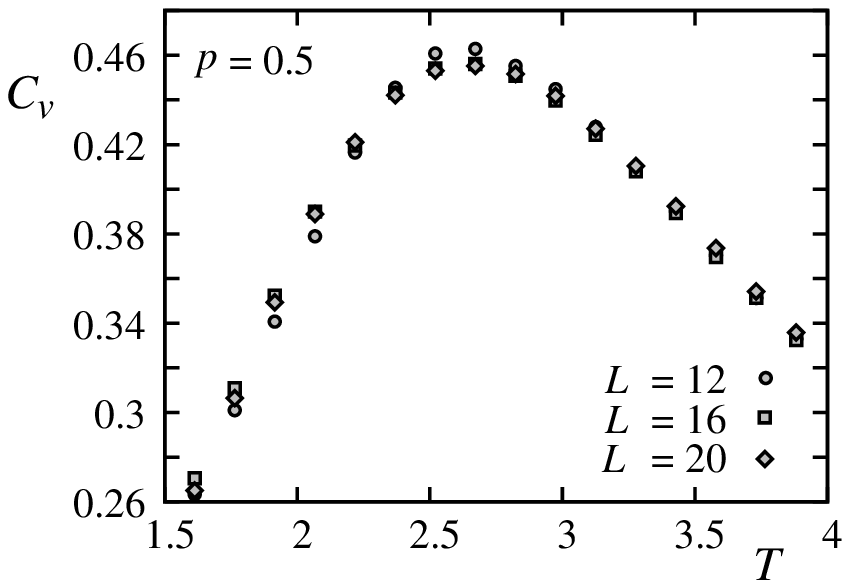,width=6cm}}
\centerline{\epsfig{file=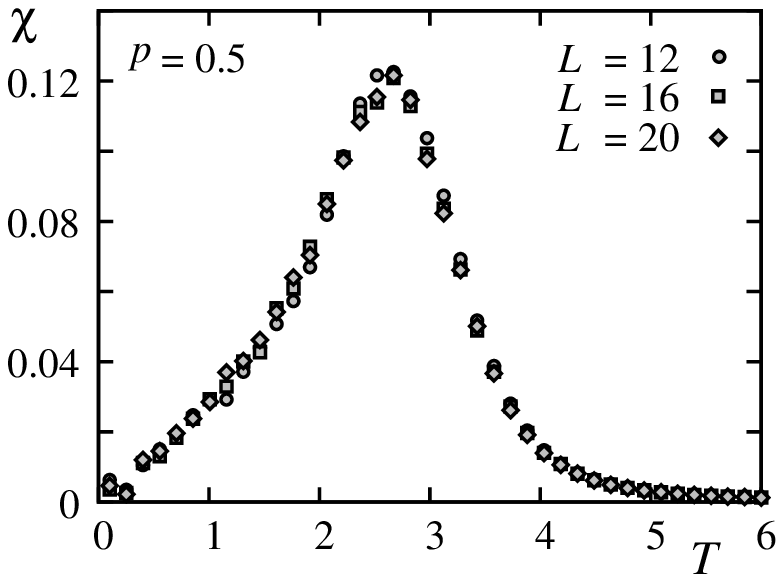,width=6cm}}
\caption{The size dependence of the specific heat (top) and the susceptibility (bottom) with $p=0.5$ for $L=12,\ 16,\ 20$.}
\label{fig:cosp0005}
\end{figure}

To close this section, let us show the size dependence for some quantities.
Figure \ref{fig:cosp0005}  shows the specific heat and the susceptibility versus $T$ with system sizes $N=4\times L^3$ where $L=12$, $16$ and $20$, for $p=0.5$.

 We  see that the size effect is not strong at the strongest disorder $p=0.5$. Our results for the SER shown above  with $p=0.5$ and $L=12$ will not be altered with larger sizes.  Of course, if we wish to calculate critical exponents we need to make a careful finite-size analysis, but this is left for future investigations.

\section{Concluding Remarks}

In this paper, we obtained three striking results:

(i) the disappearance of the latent heat when a very small amount of bond disorder ($\simeq 1\%$) is introduced into our system.  This result is in agreement with theories \cite{Aizenman,Hui,Cardy,Berker1996} and simulations on 3D models\cite{Berker2012,Fernandez2012}
having a first-order transition in their pure state.

(ii) the long-range ordering of the pure system considered here (FCC antiferromagnet) is lost at a very small amount of bond disorder ($\simeq 5\%$). This means that  the entropy created by the so-called ``order by disorder" which is at the origin of the long-range ordering in the FCC antiferromagnet is so small that it is overwhelmed by just a small bond disorder.

(iii) the SER is verified with $b=1/3$ at the freezing transition in agreement with MC simulations on other Ising SG\cite{Ogielski,Campbell2011} and with theories\cite{DeDominicis,Almeida}. This value is close to the ones experimentally observed in various SG \cite{Chamberlin,Hoogerbeets,Suzuki,Malinowski2011}. We believe that the value  $b=1/3$ at $T_c$ in the SER is universal for different kinds of SG.

Finally, we note that we studied here a dozen bond configurations of size $24\times 12^3$.  At this large number of bonds,  statistical fluctuations between samples are almost zero. We did not study therefore the size effect on the relaxation time in the present paper, reserving a priority for simulations using the whole range of ferromagnetic bond concentration which took  an enormous CPU time. In addition, we did not study SG exponents such as $\nu$ and $z$ (dynamic exponent) so the finite-size scaling is not necessary at this stage. We believe however that our results on the phase diagram and the relaxation time remain valid for larger sizes. Size effects in SG are certainly much smaller than in pure systems because of  the existence of different correlation lengths at the glass transition due to different degrees of local disorders, in the same manner as the existence of  many different exponential relaxation times whose superposition leads to the slow SER.  It is however very interesting to study the finite-size scaling to calculate the universality class of a system having a first-order transition after its crossover to a second-order due to the introduction of a random bond disorder.  This problem is left for a future study.

\section*{Acknowledgments}

The authors are grateful to A. Nihat Berker and H. Orland for helpful comments and suggestions.

V. Thanh Ngo would like to thank  the University of Cergy-Pontoise for a financial support during the course of this work. He is grateful to Vietnam National Foundation for Science and Technology Development  (Nafosted)  for support (Grant No. 103.02-2011.55).

{}

\end{document}